\documentclass[aps,prd,12pt,amssymb]{revtex4}

\begin{document}

\baselineskip=0.60cm

\newcommand{\ini}{\begin{equation}}
\newcommand{\fin}{\end{equation}}
\newcommand{\inir}{\begin{eqnarray}}
\newcommand{\finr}{\end{eqnarray}}
\newcommand{\inif}{\begin{figure}}
\newcommand{\finf}{\end{figure}}
\newcommand{\bc}{\begin{center}}
\newcommand{\ec}{\end{center}}

\def\ol{\overline}
\def\pa{\partial}
\def\ra{\rightarrow}
\def\ts{\times}
\def\df{\dotfill}
\def\bs{\backslash}
\def\dg{\dagger}

$~$

\hfill DSF-44/2003

\hfill CERN-TH-2004-015

\vspace{1 cm}

\title{Bounds for neutrinoless double beta decay in SO(10) inspired see-saw models}

\author{F. Buccella}
\author{D. Falcone}

\affiliation{Dipartimento di Scienze Fisiche, Universit\`a di Napoli, Via Cintia, Napoli,
Italy, and INFN, Sezione di Napoli}

\begin{abstract}
\vspace{1cm} \noindent By requiring the lower limit for the
lightest right-handed neutrino mass, obtained in the baryogenesis
from leptogenesis scenario, and a Dirac neutrino mass matrix
similar to the up-quark mass matrix we predict small values for
the $\nu_e$ mass and for the matrix element $m_{ee}$ responsible
of the neutrinoless double beta decay, $m_{\nu_e}$ around $5\cdot10^{-3}$ eV
 and $m_{ee}$ smaller than $ 10^{-3}$ eV,
respectively. The allowed range for the mass of the heaviest
right-handed neutrino is centered around the value of the scale of
B - L breaking in the SO(10) gauge theory with Pati-Salam
intermediate symmetry.
\end{abstract}

\maketitle

\newpage

The robust experimental evidence \cite{nu1,nu2} for neutrino
oscillations \cite{p}, with increasing precision on the square
mass differences and mixing angles \ini \Delta m^2_s=7 \cdot10
^{-5} eV^2 \fin \ini (\tan{\theta_s})^2 = 0.4 \fin \ini \Delta
m^2_a=2.6 \cdot10^{-3}eV^2 \fin \ini(\tan{\theta_a})^2=1 \fin should
be a milestone for  the extension of the standard model. Indeed
the smallness of neutrino masses points in the direction of
theories like the unified SO(10) model \cite{soio}, where the
see-saw formula \cite{seesaw} \ini m_L = -m_D^T M_R^{-1} m_D \fin
arises naturally with the elements of the Dirac mass matrix of the
order of the charged fermion masses and the elements of the
Majorana mass matrix at the scale of the spontaneous symmetry
breaking of $B-L$. Right-handed neutrinos of high mass have been
advocated in the baryogenesis from leptogenesis scenario \cite{fy}
and a lower limit for the mass of the lightest neutrino around $5
\cdot 10^8$ GeV has been found \cite{di}.

In a recent paper Akhmedov, Frigerio and Smirnov \cite{afs}
studied the compatibility of the seesaw mechanism with what is
known on neutrino oscillations and the leptogenesis scenario, and
reached the conclusion that the two lightest states should be
almost degenerate. Here we study the consistency of the see-saw
formula with the leptogenesis framework within the hypothesis that
$m_D$ has strong similarities with $m_u$, as expected in SO(10)
theories. We do not require strict equality (which would follow
from the assumption that the electroweak Higgs transforms as 10
representations), since it does not hold for $m_d$ and $m_e$, but
assume, as in \cite{afs}, the eigenvalues of $m_D$ to be 1 MeV,
400 MeV and 100 GeV, respectively, with the same hierarchical
pattern as the up quarks, and small mixing angles for the matrices
$V_R$ and $V_L$ which diagonalize $m_D$ in the basis where $m_e$
is diagonal: \ini m_D = V_R~ \text{diag} (m_D)~ V_L^{\dg}. \fin
We define \ini
M_h=(\text{diag}(m_D))_{hh} \fin and $m_i$ (i = 1,2,3) the eigenvalues
of $m_L$.
One has for the mass of the lightest neutrino the upper limit \ini
|m_1|^3 <
|m_1m_2m_3|=
\frac{(M_1M_2M_3)^2}{M^R_1M^R_2M^R_3}
< (0.23~ \text{eV})^3 \fin comparable with the limit coming from astrophysics
 \cite{s}.
 From Eq.(6) and the inverse see-saw formula \ini
M_R= -m_D m_L^{-1} m_D^T \fin it is easy to derive \ini M^R_{ab}=
-V^R_{af} V^R_{bg} M_f M_g A^L_{fg} \fin where 
 \ini A^L_{fg}=(m_L^{-1})_{cd}
V^{*L}_{cf} V^{*L}_{dg} \fin which shows the intriguing property
that the part proportional to $M_f M_g$ of all the matrix elements
of $M_R$ is also proportional to the same factor  $A^L_{fg}$. This
part may be diagonalized by the transformation induced by any
unitary matrix with $(V^R_{13}, V^R_{23}, V^R_{33})$ in the last
line.

If we take $\theta_{atm}=45^o$, $\theta_{13}=0^o$ and  define $c =
\cos{\theta_s}$, $s = \sin{\theta_s}$, we get the following
elements of the effective neutrino mass matrix:
\ini
m_{ee}=c^2 m_1+s^2 m_2
\fin
\ini
m_{e \mu}=cs(m_1-m_2)/\sqrt2
\fin
\ini
m_{e \tau}=cs(m_1-m_2)/\sqrt2
\fin
\ini
m_{\mu \mu}=(s^2 m_1+c^2 m_2+m_3)/2
\fin
\ini
m_{\mu \tau}=(s^2 m_1+c^2 m_2-m_3)/2
\fin
\ini
m_{\tau \tau}=(s^2 m_1+c^2 m_2+m_3)/2,
\fin
and a similar expression for $m_L^{-1}$ with the exchange $m_i \ra
1/m_i$. From the expression for $m_L^{-1}$ we get
$$
A^L_{33}=[c V^{L*}_{13}+\frac{s}{\sqrt2}(V^{L*}_{23}+V^{L*}_{33})]^2
\frac{1}{m_1}+
[s V^{L*}_{13}-\frac{c}{\sqrt2}(V^{L*}_{23}+V^{L*}_{33})]^2 \frac{1}{m_2}+
\frac{1}{2}[V^{L*}_{23}-V^{L*}_{33}]^2 \frac{1}{m_3}=
$$
\ini \frac{1}{2} [V^{L*}_{33}]^2 (\frac{s^2
}{m_1}+\frac{c^2}{m_2}+\frac{1}{m_3})  + ... \fin where we have
selected the term proportional to $(V^{L*}_{33})^2$, since we
assume that, as it happens in the CKM matrix \cite{ckm}, the
largest matrix elements are the diagonal ones. From now on we
shall take $\tan^2 \theta_s = 0.4$ \cite{nu1}.

If $|m_1| \ll (\tan^2 \theta_s) |m_2|$, the lightest eigenvalue of
$M_R$ would be $M^R_3 \simeq M_3^2/7 m_1$ and then the product
\ini M^R_2 M^R_1 \simeq \frac{7 (M_2 M_1)^2}{ |m_2 m_3|} \sim
2.6 \cdot 10^{15} ~\text{GeV}^2, \fin which implies $M^R_1 < 5.1
\cdot 10^7$ GeV, an order of magnitude smaller than the lower
limit required in the leptogenesis scenario.

To get some cancellation in the r.h.s. of Eq.(18), $|m_1|$ should be
at least $\simeq (2/5) |m_2|$. This would imply for the product of
the eigenvalues of $M_R$ \ini M^R_3 M^R_2 M^R_1 \le
\frac{21(M_3 M_2 M_1)^2}{10\Delta m_s^2 \sqrt{\Delta
m_a^2}} \simeq 10^{30} \text{GeV}^3, \fin and therefore, in order
to be $M_1^R \ge 5 \cdot 10^8$ GeV, one should have
\ini
 M_3^R \le 4 \cdot
10^{12} GeV
\fin
 which requires, according to Eq.(18), \ini
|\frac{s^2}{m_1}+ \frac{c^2}{m_2}+ \frac{1}{m_3}| < 0.8 eV^{-1}
\fin and a cancellation between the three contributions, in fact
$\frac{1}{|m_3|} > 4$ eV$^{-1}$. We should also have \ini M^R_3 M^R_2
\le 2 \cdot 10^{21} \text{GeV}^2 \fin which implies that the
coefficient of $(M_3 M_2)^2$ in the expression of $M^R_3
M^R_2$ be less than 1.25 eV$^{-2}$.

For hermitean matrices, if the lightest eigenvalue is smaller than
the other two, the product of them is approximately the sum of the
minors corresponding to the three diagonal matrix elements. Let us
consider then \ini M^R_{aa} M^R_{bb}-(M^R_{ab})^2=
\frac{\epsilon_{abz}^2 \epsilon_{rsp} \epsilon_{tnq}}{4 (m_1 m_2
m_3)} M_r M_s M_t M_n V^R_{pz} V^R_{qz}
\tilde{A^L_{pq}}= \epsilon^2_{abz}(V^R_{1z})^2\frac{M^2_2
M^2_3}{m_1 m_2 m_3}\tilde{A^L_{11}}+... \fin where
$\tilde{A^L_{pq}}= V^L_{ip}  V^L_{jq} m^L_{ij}$, which shows that
the terms proportional to $(M_3 M_2)^2$ for all the minors
considered are multiplied by the factor
$$
|\frac{1}{m_2 m_3}[c V^L_{11}+\frac{s}{\sqrt2}(V^L_{21}+V^L_{31})]^2+
$$
$$
\frac{1}{m_1 m_3}[s V^L_{11}-\frac{c}{\sqrt2}(V^L_{21}+V^L_{31})]^2+
$$
\ini
\frac{1}{2m_1 m_2}[(V^L_{21}-V^L_{31})]^2| < 1.25~\text{eV}^{-2}.
\fin
Since we have \ini \frac{1}{|m_2 m_3|}\ge 25~\text{eV}^{-2}, \fin to get
the coefficient of $(M_3 M_2)^2$ smaller than $1.25~\text{eV}^{-2}$, we
need a cancellation between the three terms in Eq.(25). Let us
see if it is possible to obey at the same time the limits given by
Eqs.(22) and (25) with $m_L$ given by Eqs. (13-18) and
the form
\ini
V^L=\left( \begin{array}{ccc}
\cos \rho & \sin \rho & 0 \\
-\sin \rho & \cos \rho & 0 \\
0 & 0 & 1
\end{array} \right)
\fin which looks like the CKM matrix up the order $\lambda$.
Eq.(25) takes the form 
$$ |\frac{1}{m_2 m_3}(c^2 (\cos{\rho})^2
+ s^2 (\sin{\rho})^2 - \sqrt{2} c s \cos{\rho} \sin{\rho}) +
$$
$$
\frac{1}{m_1 m_2}(s^2 (\cos{\rho})^2 + \frac{c^2}{2}
(\sin{\rho})^2 + \sqrt{2} c s \cos{\rho} \sin{\rho}) +
$$
\ini\frac{1}{m_1 m_2} \frac{(\sin{\rho})^2}{2}| < 1.25
~(\text{eV})^{-2} \fin
In \cite{pw} one found the solutions of Eq.(22) with a vanishing
r.h.s. consistent with \ini m^2_2 - m^2_1 = \Delta m^2_s = 7 \cdot
10^{-5} ~\text{eV}^2 \fin \ini m^2_3 - c^2 m^2_2 - s^2 m^2_1 = 2.6
\cdot 10^{-3} ~\text{eV}^2, \fin one with the same sign for $m_2$ and
$m_3$ and \cite{BF} 
\ini 
\frac{m_1}{m_2} =-(\tan{\theta_s})^2-\frac{m_1}{m_3 c^2}\simeq-
\frac{s^2}{c^2+0.16},
\fin
the other with the same sign for $m_1$ and $m_3$, $m_3$ larger
than the other two, and \ini m_1 \simeq - m_2, \fin  \ini m_1
\simeq \cos{2 \theta_s} m_3 = \frac{3}{7} m_3.
\fin
These two solutions are slightly
modified by the small upper bound in the r.h.s. of Eq.(22) into
\ini
\frac{m_1}{m_2} \simeq - (\tan{\theta_s})^2\frac{c^2 \pm 0.8 \cdot eV^{-1}
\sqrt{\Delta m^2_s}}{c^2 + \sqrt{\frac{\Delta m^2_s}{\Delta m^2_a}}} =
\frac{-2 \pm 0.02}{6.12}
\fin
\ini
\frac{m_1}{m_3} \simeq \cos{2 \theta_s} (1 \pm 0.8 \cdot eV^{-1} \sqrt{\Delta m^2_a}) =
\frac{3}{7} (1 \pm 0.04).
\fin
There are also 
solutions, with $m_1$ and $m_2$ with the same sign and $|m_1|>
0.11$ eV or $ > 2.5$ eV (which is excluded by astrophysics \cite{s}).
However only the solution of Eq.(22) with the same sign
for $m_2$ and $m_3$ satisfies Eq.(28) with a small value of
$\rho$.
By considering only the term proportional to $(V^{L*}_{33})^2$ in 
Eq.(19), one neglects terms at the order \ini
\frac{4 \sqrt{2} c V^{L*}_{13}}{s V^{*L}_{33}} 
\simeq 9 \frac{V^{L*}_{13}}{V^{L*}_{33}}
\fin
for the relative coefficients of $\frac{1}{m_1}$ and $\frac{1}{m_2}$
and
\ini
\frac{4 V^{L*}_{23}}{V^{L*}_{33}}
\fin
for the relative coefficients of $\frac{1}{m_1}$ and $\frac{1}{m_3}$.
Should one take for the ratios of the matrix elements of $V^L$ the 
corresponding values of the CKM matrix, these uncertainties would be
10\% and 100\% and would effect the first and the second term in the
denominator of the r.h.s. of Eq.(31), respectively.
By taking into account all the uncertainties one has
\ini
-0.45 < \frac{m_1}{m_2} < -0.25
\fin
which implies
\ini
\-4.23 \cdot 10^{-3}< m_1 < -2.16 \cdot 10^{-3}~\text{eV} \fin \ini
8.64 \cdot 10^{-3} < m_2 < 9.4 \cdot 10^{-3}~\text{eV}
\fin \ini 5.15 \cdot 10^{-2} < m_3 < 5.18 \cdot 10^{-2}~{eV}. \fin 
We predict a small
value for the neutrinoless double beta decay parameter \ini
M_{ee}=\sin^2 \theta_s m_2 +\cos^2 \theta_s m_1 =-m_1 m_2/m_3 
\pm 0.8 \text{eV} ^{-1} m_1 m_2
\fin
which implies $M_{ee}$ to be in the range $ (3.5 - 8)
\cdot 10^{-4} \text{eV}$ \ and a slightly larger value for the
effective electron neutrino mass \ini m_{\nu_e} =\sin^2 \theta_s
|m_2| +\cos^2 \theta_s |m_1|
\fin
in the range $(4-5.6) \cdot 10^{-3} \text{eV}$.
For simplicity we have considered real $m_i$'s, but our results
would not change by considering complex $m_i$'s.

By taking
$A^L_{33}=0$ and $\tilde{A^L_{11}}=0$ one gets \ini \tan{\rho} =
-\frac{\sqrt{2} s c (m_2- m_1)}{m_3 + c^2 m_2 + s^2 m_1} 
\fin in the range $(-0.15, -0.12)$ about twice smaller than $\tan{\theta_c}$.

>From the identity
\ini A^L_{33} A^L_{22}- (A^L_{23})^2 = \tilde{A}^L_{11} / m_1 m_2 m_3
\fin it follows that, if we ask that no term proportional to
$M_3^2$ is present in $M_R$ and no term proportional to
$(M_3 M_2)^2$ is contained in the diagonal minors of $M_R$,
there is no term proportional to $(M_3 M_2)^2$ in $M_R$. In
conclusion, the largest contribution to $M_R$ would be the ones
proportional to $M_3 M_1$ and $M_2^2$ which are
multiplied by the factors $A^L_{31}$ and $A^L_{22}$, respectively:
\ini M^R_{ab}= -M_3 M_1 V^R_{a1} V^R_{b3} A^L_{31} -
M_2^2 V^R_{a2} V^R_{b2} A^L_{22} + ... \fin 
where we have not written
the terms proportional to $M_2 M_1$ and $M_1^2$. In
general \ini \text{Det} M^R= - (M_3 M_2 M_1)^2
\text{Det} A^L \fin and therefore \ini \text{Det} A^L =(m_1 m_2
m_3)^{-1}. \fin
For real $M_R$, which is not what we want to get
the CP violation needed to produce baryogenesis, but is a property
of the CKM matrix at the first order in $\sin \theta_c$, we have
the eigenvalues of $M_R$ as $- M_2^2 A^L_{22}$ and $\pm (M_3
M_1)/\sqrt{A^L_{22} m_3 m_2 m_1}$. Since \ini A^L_{22}= \sin^2
\rho (\frac{c^2}{m_1}+\frac{s^2}{m_2}), \fin the eigenvalues are in
the ranges $(5-6.8) \cdot 10^8$ GeV and $\pm (4-5) \cdot 10^{10}$ GeV, 
respectively. The lightest
right-handed neutrino is just above the lower limit $5 \cdot 10^8$
GeV from leptogenesis. We have deduced the consequences of
$A^L_{33}=0$ and $\tilde{A^L_{11}}=0$ to display an example with
the lightest right-handed neutrino mass larger than $5 \cdot 10^8$
GeV.

In general the mass of the heaviest right-handed neutrino
will be in the range $10^{10}$ - $5\cdot 10^{12}$ GeV with the
geometrical mean of the two extrema very near to the scale of $B-L$
symmetry breaking in the $SO(10)$ unified theory  with Pati-Salam
\cite{ps} intermediate symmetry, $3 \cdot10^{11}$ GeV \cite{aa}.

Our assumption of small mixing values for $V^L$ and on the value of 
the product of the two smallest eigenvalues of $m_D$ ( $4 \cdot10^2
MeV^2$ ) excludes the high values of $M^R_3$ proposed in the literature
\cite{BS}.

The upper limit given by Eq.(21) implies a small coefficient for the 
contribution proportional to $M^2_3$ to the matrix elements of $M^R$,
which is welcome to get the lepton asymmetry needed for leptogenesis,
which would vanish, should one neglect the other contributions.

We want to stress that the small values
found for $A^L_{33}$ and $\tilde{A^L_{11}}$ are not unnatural
tunings; just the opposite: since Eq.(5) gives $m_L$ in terms of
$M^R$ (and of $m^D$), the fact that the  matrix
elements of $M^R$ ( as well as its eigenvalues ) are not so
different, as it happens for $m_L$, makes the choice described
here a natural one.

$~$

Aknowledgements.

One of us (FB) aknowledges a clarifying discussion with Prof. Ferruccio
Feruglio and the hospitality at the theory division of CERN, where part of this work
has been done.


\begin{thebibliography}{199}

\newpage

\bibitem{nu1} Q.R. Ahmad et al. (SNO Collaboration), {\it Phys. Rev. Lett.}
{\bf 89} (2002) 011301; 011302

K. Eguchi et al. (KamLAND Collaboration), {\it Phys. Rev. Lett.}
{\bf 90} (2003) 021802

\bibitem{nu2} Y. Fukuda et al. (SuperKamiokande Collaboration), {\it Phys. Rev. Lett.}
{\bf 81} (1998) 1562

\bibitem{p} B. Pontecorvo, {\it Sov. Phys. JETP} {\bf 6} (1958) 429; {\bf 7} (1958) 173

\bibitem{soio} H.Georgi, in Particles and Fields, ed. C. Carlson (AIP), 1975;

H. Fritzsch and P. Minkowski, {\it Ann. Phys.} {\bf 93} (1975) 73

\bibitem{seesaw} M. Gell-Mann, P. Ramond and Slansky in
Supergravity, eds. P. von Nieuwenhuizen and D. Freedman
(North-Holland, Amsterdam, 1979) ;

T. Yanagida in United Theories and Baryon
Number in the Universe, eds. O. Sawada and A. Sugamoto (KEK,
Tsukuba, 1979);

R, N, Mohapatra and G. Semjanovic, {\it Phys. Rev. Lett.} {\bf 44} (1980) 912

\bibitem{fy} M. Fukugita and T. Yanagida, {\it Phys. Lett.} {\bf B 174} (1986) 45

\bibitem{di} S. Davidson and A. Ibarra, {\it Phys. Lett.} {\bf B 535} (2002) 25;

W. Buchmuller, P. Di Bari and M. Plumacher, {\it Phys. Lett.}
{\bf B 547} (2002) 128

\bibitem{afs} E.Kh. Akhmedov, M. Frigerio and A.Yu. Smirnov, {\it J. High Energy Phys.}
{\bf 09} (2003) 021

\bibitem{s} D. N. Spergel et al., {\it Astrophys. J. Suppl.} {\bf 148} (2003) 175

\bibitem{ckm} N. Cabibbo, {\it Phys. Rev. Lett.} {\bf 10} (1963) 531;

M. Kobayashi and T. Maskawa, {\it Prog. Theor. Phys.} {\bf 49} (1973) 652

\bibitem{pw} M. Abud and F. Buccella, {\it Int. J. Mod. Phys.} {\bf A 16} (2001) 609;

D. Falcone, {\it Phys. Rev.} {\bf D 66} (2002) 053001

\bibitem{BF} F. Buccella and D. Falcone, {\it Mod. Phys. Lett.} {\bf A18} (2003) 1819

\bibitem{ps} J. C. Pati and A. Salam, {\it Phys. Rev.} {\bf D10} (1074) 275

\bibitem{aa} D. Lee, R. N. Mohapatra, M. K Parida and M. Rani,
{\it Phys. Rev.} {D 47} (1993) 264;

B. Ananthanarayan, K. S. Babu and Q. Shafi, {\it Nucl. Phys.} {\bf B 428} (1994) 19;

F.Acampora, G. Amelino-Camelia, F. Buccella, O. Pisanti, L. Rosa and
T. Tuzi, {\it Nuovo Cim.} {\bf A 108} (1995) 37

\bibitem{BS} G. C. Branco, R. Gonzales Felipe, F. R. Joachim, I. Masina, M. N. Rebelo
and C. A. Savoy, {\it Phys. Rev.} {\bf D67} (2003) 073025;

S. Pascoli, S. T. Petcov and W. Rodejohann, {\it Phys. Rev.} {\bf D68} (2003) 093007




\newpage

\end{thebibliography}
\end{document}